# Ruthenium Tetrazole-Based Electroluminescent Device: The Key Role of Counter-Ions for the Light Emission Properties

Hashem Shahroosvand[a,*], Leyla Najafi[a,b,c], Ahmad Sousaraei[a], Ezeddin Mohajerani[d], Mohammad Janghouri[d], and Francesco Bonaccorso[b,*]

[a] Chemistry Department, University of Zanjan, Zanjan, Iran
[b] Istituto Italiano di Tecnologia, GrapheneLabs, 16163 Genova, Italy
[c] Dipartimento di Chimica e Chimica Industriale, Università degli Studi di Genova, Via Dodecaneso 31, Genoa, 16146, Italy
[d] Laser and Plasma Research Institute, Shahid Beheshti University, Tehran, Iran

* Corresponding authors: Hashem Shahroosvand, shahroos@znu.ac.ir; Francesco Bonaccorso, francesco.bonaccorso@iit.it.

## Abstract

Light-emitting electrochemical cells (LECs), thanks to their simple device structure and the tunable emission wavelength of the light-emitting layer, are emerging as a class of electrochemical device candidate for the development of next-generation solid-state lighting. The possibility to tune on-demand the energy levels of Ruthenium(II) polypyridyl complexes, makes them ideal candidates as light-emitting layer. However, the optimization of the latter is not trivial and several issues, such as charge injection and electron and hole transport, have still to be solved to enhance the LEC performances. Here, we demonstrate how the exploitation of small counter anion ($BF_4^-$) enhances the light emission performance of ruthenium tetrazole complexes in light-emitting diodes. In comparison with neutral ruthenium tetrazole complexes, cationic Ru tetrazole ones, containing $BF_4^-$ ion, show a reduction of turn-on voltage from 7 to 5 V, at high luminous efficiency (1.49 cd/A) and applied voltage of 12 V, and a twofold improvement of the luminance. Moreover, complexes containing $BF_4^-$ counter ion show better stability of luminance over time than other complexes without counter ion.

## Introduction

The quest for next-generation materials for the development of highly efficient and environmentally friendly optoelectronic devices is the most demanding challenge for material scientists.[1-3] Light-emitting diodes (LEDs) and organic light-emitting diodes (OLEDs) are two main electroluminescent devices being developed in solid-state lighting (SSL). The latter represents an important branch in modern optoelectronic industry based on semiconductor emitters relying on both organic and inorganic materials that have enabled the realization of bright and energy-saving light sources.[4] In particular, organic electroluminescence (EL) is the electrically driven emission of light from organic and/or inorganic materials. Generally, OLEDs consist of one organic compound as an emitter layer sandwiched between two metallic layers (Figure 1). In fact, the figures of merit (FoM) of OLEDs are influenced by different factors, including electrodes material, emitter, hole- and electron- transport layer, and polymer host, as well as transport layers thickness, purity of emitter and contamination.[5] The most important parameters to be tuned for the optimization of OLED performances are the balance in terms of density and mobility of charge carriers, and the matching of energy levels of the highest occupied molecular orbital (HOMO) and the lowest unoccupied molecular orbital (LUMO) of light emitting layer, hole and electron transport layers.[6,7] The progress of OLED technology, mainly driven by the search for high performance OLED material, was initiated in 1972, when Tokel and Bard exploited the $Ru(bpy)_3^{2+}$ complex as emitter.[8]

After many years of development following the Bard's pivotal work,[8] J. K. Lee et al., demonstrated that Ru polypyridyl complexes show luminance as high as 100 cd/m$^2$ at low turn-on voltage, i.e., in the 2.5-3.5V range.[9]

Light-emitting electrochemical cells (LECs) are another class of electroluminescent devices candidate for the development of next-generation SSLs.[10] A LEC consists of an ionic luminescent material, sandwiched between two electrodes, i.e., anode usually made of indium tin oxide (ITO)and cathode based on metals such as aluminum or gold in an ionic environment.[11] The luminescent material is either an ionic transition-metal complex (iTMC) or conjugated light-emitting polymer.[12,13] Light-emitting electrochemical cells are characterized by: (1) turn on voltage close to the optical band gap of the light emitting layer (LEL); (2) weakly dependence of the FoM on the emitter thickness; (3) symmetrical current-voltage characteristic; (4) external quantum efficiency (EQE) independent on the electrode work-function.[14,15] The LEC working principle is based on the reduction of the injection barrier for holes and electrons due to the separation of the counter ions in the LEL upon bias application.[16] Therefore, LEC can be considered as a single-layer version of OLED, *i.e.*, one opto-electronically active layer with efficient light emission.[17] Due to the device structure simplicity, LEC can be easily prepared via solution-processing technologies such as spin-coating[18,19] and screen printing,[20] just to cite a few. Thus, LEC are emerging as a promising alternative to OLEDs. In fact, OLEDs suffer high manufacturing cost, due to the multilayer evaporation process, and the need for technological encapsulation techniques to preserve both the devices opto-electronic properties and lifetime.[21] Although the device structure is quite simple, the optimization of the light-emitting layer is not trivial.[22]

The crucial point for device optimization relies on the emission wavelength tuning of the emitters, across the visible region, to produce a wide range of colors, in particular, white for display applications.[23,24] Iridium cyclometalates are the benchmark complexes.[25,26] However, the demonstration of multiple colors is a critical point for this class of emitters, which typically emit in the blue region.[27] Moreover, iridium is expensive, posing major challenges toward the commercialization of OLED technologies based on such emitters. Therefore, less expensive metals, such as ruthenium, are of great interest for the development of OLEDs.[28]

Novel ruthenium compounds could be interesting materials for OLEDs,[28,29] due to emission colors tuning, especially in the green and red regions of the electromagnetic spectrum, which can be controlled with the introduction of different ligands into the ruthenium complexes.[27] Similarly to other electroluminescent device,[30] ionic Ru(II) polypyridyl complexes have been exploited in LECs due to tunable emission wavelengths guaranteed by the tuning of their HOMO‑LUMO energy levels.[31-33] Moreover, by means of functionalization (*i.e.*, attachment of functional groups to the periphery of the complexes) it is possible to further expands the aromaticity of the Ru polypyridyl complexes[34] (see Table S1 in the supplementary information-S.I.-). Tetrazoles have been used as precursors of different N- donor ligands,[35] finding applications in (opto)electronic and information recording devices.[36]

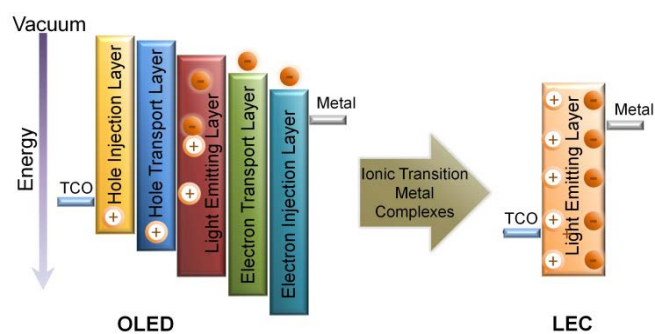

Figure 1 Schemes of a typical OLED (left) and LEC (right).

Thus, the incorporation of these ligands in one hybrid-emitting structure may provide another pathway towards the optimization of tunable color emission in electroluminescence devices. However, several issues have to be solved to optimize LEC performances. In particular, the presence of counter ion in iTMC plays a key role in promoting the charge injection from the cathode, as well as the electron and hole transport through the Ru tetrazole LEC device. Furthermore, the charge transport at the anode is strongly influenced by the physical properties of the counter-ion of ionic transition metal complexes.[8,14] For example, small size anions (*e.g.*, $BF_4^-$) can enhance the charge carriers transport for a fast recombination with the opposite carrier, creating an excited Ru complexes.[8] Contrary, large size anions (*e.g.*,$PF_6$) produce a slow charge separation.[14] Therefore, the search for the "ideal" counter ion that guarantee a balanced injection of negative and positive charges across the LEC and the recombination in the light-emitting layers is still a fervent research area.

Here, we present the optoelectronic performances of six Ru polypyridyl emitters, as single active components in OLEDs, containing the 2-(5-(pyridin-2-yl)-2H-tetrazole-2-yl) acetic acid(PTA) as a fundamental ligand, with 2, 2-bipyridine (bpy), 1, 10- phenanthroline (phen) and 2- pyridine (1H-tetrazole-5-yl) (pyTz) as ancillary ligands. We show how cationic Ru tetrazole complexes containing $BF_4^-$ ion show better optoelectronic performances, with respect to neutral ruthenium tetrazole complexes. In fact, we demonstrate a reduction of turn-on voltage from 7 to 5 V, at high luminous efficiency (1.49 cd/A) and applied voltage of 12 V, together with a twofold improvement of the luminance, with respect to that of the Ru tetrazole derivatives.

Overall, with the present work we show the importance of (i) the substitution on the Tz moiety in the pyTz ligand, (ii) the key role of counter ion to increase the LEC performance and stability, (iii) the occurrence of Förster transfer mechanism to reach the maximum charge transfer in the active layer.

## EXPERIMENTAL

We synthesized and characterized six Ru complexes, named: [Ru(PTA)(phen)(SCN)$_2$] (1), [Ru(PTA)(phen)$_2$](BF$_4$)$_2$ (2), [Ru(PTA)(bpy)(SCN)$_2$] (3), [Ru(PTA)$_2$(phen)](BF$_4$)$_2$(4), [Ru(PTA)2(bpy)](BF$_4$)$_2$ (5), [Ru(PTA)(pyTz)(bpy)]BF$_4$ (6), respectively, (Please see electronic supporting information -ESI- for more details concerning the synthesis and characterization of the six Ru complexes). The molecular structure of complex (6) and the HOMO and LUMO electron densities are shown in Figure 2.

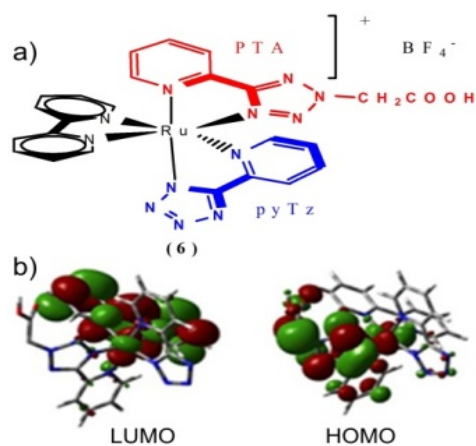

**Figure 2**. a) Chemical structures of complex (6). b) Electron density contours for the HOMO and LUMO of complex (6) obtained from density functional theory (DFT) method through B$_3$LYP/LANL2DZ basic set.

## RESULT AND DISCUSSION

### Photo-physical and electro-chemical studies

The Ru(PTA) complexes emit in the yellow-red region of the electromagnetic spectrum at room temperature, with the maximum photoluminescence (PL)wavelength ($\lambda_{max}$) in the 560-615 nm region. In all Ru(PTA) cases, the PL emission originates from $^3$MLCT states involving the charge transfer from polypyridyl and tetrazole ligands to t$_{2g}$ of ruthenium center, as demonstrated by the shape and position of the emission band in the 550- 700 nm region, and by the lifetime values, which are in the expected range for Ru–tetrazole complexes, *e.g.*, 100-500 ns in degassed acetonitrile.[37] The PL emission quantum yields are calculated by comparison with the known [Ru(bpy)$_3$]$^{2+}$ one ( $\phi_{std}$ = 9.5 %) in acetonitrile solution at room temperature with excitation wavelength of 405 nm (see Figure S 1).[38,39] The absorption spectra of the Ru(PTA) complexes, measured in acetonitrile solution at room temperature are shown in Figure 3 and their data are summarized in table 1. (Please see ESI,S1 for further information).

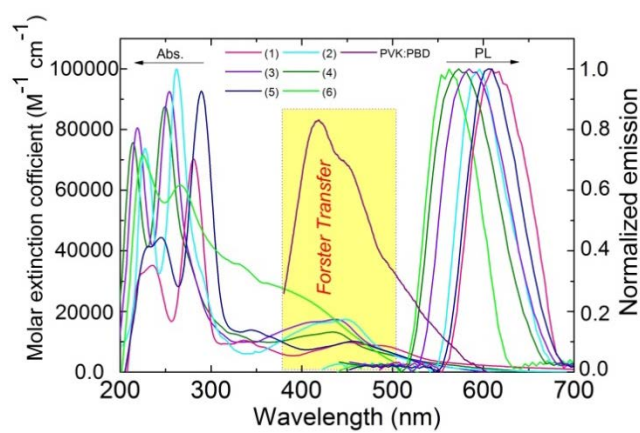

**Figure 3**. Absorption and PL ( $\lambda_{exc}$=405 nm) spectra of Ru(PTA) complexes (1-6) in acetonitrile solution 10$^{-5}$M. The PL spectrum of PVK:PBD is also reported.

**Table 1**. Photo-physical properties of Ru(PTA)(1-6) complexes, $^a$in air-equilibrated acetonitrile and $^b$in degassed acetonitrile. $\lambda_{max}$ is the maximum emission wavelength, $\Phi$ is the quantum yield, while $\tau$ is the lifetime.

|     | Absorption(nm)          |              | $\lambda_{max}$ (nm) | $\Phi^a$ (%) | $\tau^b$(ns) [$\pm$10] |
| --- | ----------------------- | ------------ | ------- | ----- | ---- |
|     | n, $\pi$ - $\pi$ *      | MLCT         |         |       |      |
| (1) | 226, 262                | 454, 489     | 614     | 3.2   | 160  |
| (2) | 225, 263                | 415, 448     | 597     | 5.4   | 112  |
| (3) | 228, 244, 288           | 410, 443     | 584     | 4.0   | 195  |
| (4) | 224, 259                | 407, 436     | 573     | 4.5   | 148  |
| (5) | 227, 245, 287           | 420, 451     | 606     | 7.5   | 205  |
| (6) | 232, 273                | 415(broad)   | 562     | 7.8   | 230  |

Cyclic voltammetry are carried out both to investigate the redox properties and to determine the LUMO and HOMO energy levels of the Ru(PTA) complexes.

In the region of the positive potentials, the Ru(PTA) complexes exhibit a single one-electron reversible process,[40] which is attributed to the oxidation of the Ru(II) center to Ru(III).[41] Cyclic voltammetry results show variation in the redox potential (E$_0$) of Ru(PTA) complexes with different ligand substitution (S.I. Figure S5), with values ranging from +0.88 V to +1.22V vs Ag/Ag$^+$.[37] The peaks of Ru(PTA) complexes in the −1.2 to -1.8V vs Ag/Ag$^+$ potential region can be assigned to the bpy or phen ligand, since they are easier to reduce than pyTz and PTA ligands. The addition of electron withdrawing/donating groups to the periphery of the bpy ligand shifts the position of the oxidation-reduction peaks (S.I. Figure S5).

### Electroluminescence performances

To investigate the EL behavior of the Ru(PTA) complexes, OLED devices are fabricated based on Ru(PTA)(L), where L is the ancillary ligand as the active layer. The device has the following structure: ITO/PEDOT-PSS/PVK/Ru complex/PBD/Al, see experimental and Figure S 6. In the EL spectra reported in Figure 4, Ru complexes show a significant red shift (~200nm) compared with the PVK/PBD EL spectra.

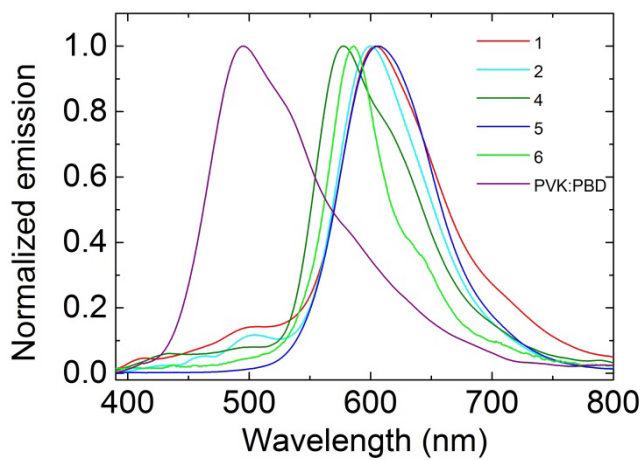

Figure 4. EL spectra of PVK:PBD and Ru complexes (1-6).

Figure 5 shows the emission colors of single-layer emitter devices based on (1-6) in a CIE chromaticity diagram (CIE=Commission Internationale de L'Eclairage; where the ideal white has x= 0.33, y= 0.33). As indicated by the EL spectra of the complexes, also the CIE figures show the influence of ancillary ligand for the color tuning.

The red shift of complexes from 568 nm to 612 nm, is consistent with energy transfer phenomenon between HOMO and LUMO in the emitting layer. The EL spectra also indicate that the pyTz ligand in [Ru(PTA)(pyTz)(bpy)]BF$_4$(6) blue shift the EL with respect to the one of [Ru(PTA)$_2$(bpy)](BF$_4$)$_2$ (5). This is attributed to the presence of N- atoms in Tz ligand, which needs higher energy than C-heterocycle as electron-deficient atoms.[42] The shift of EL emission wavelength of Ru(PTA) complexes strongly depends on the nature of the ancillary ligands coordinated to Ru(II) ion.

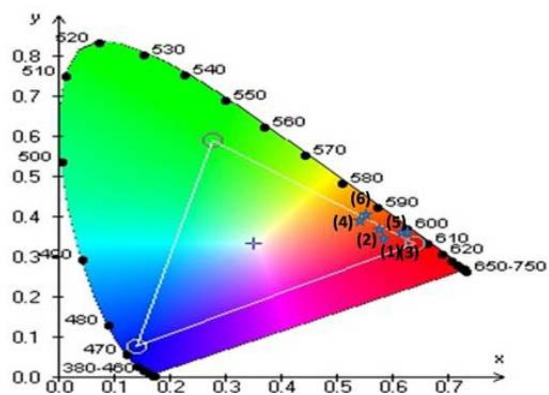

Figure 5. CIE coordinates of Ru (PTA) (1-6).

Table 2. Device characteristics of Ru(PTA) complexes(1- 6). Complex (7) is obtained from neutral Ru Tz [Ru(phen)$_2$(TzBA)$_2$] @ a=20 V; b= 16 V.[37] FWHM=Full Width at Half Maximum.

|     | EL max(nm) | CIE (x; y) | FWHM (nm) | Maximum Current density (mA· cm-2) @ 8V | Turn-on voltage (V) | Luminous efficiency (cd· A-1)@ 12V | Luminance (cd m -2 ) @ 12V | Power efficiency (lm· W-1) |
|-----|------------|------------|-----------|-----------------------------------------|---------------------|------------------------------------|----------------------------|----------------------------|
| (1) | 612 | (0.587; 0.359) | 94.2 | 50 | 7.5 | 0.60 | 510 | 0.09 |
| (2) | 590 | (0.550; 0.404) | 86.2 | 242 | 6.0 | 1.08 | 340 | 0.03 |
| (3) | 583 | (0.630; 0.365) | 86.2 | 118 | 7.5 | 0.79 | 650 | 0.11 |
| (4) | 575 | (0.630; 0.365) | 89.3 | 271 | 6.0 | 1.10 | 425 | 0.04 |
| (5) | 600 | (0.559;0.413) | 87.3 | 324 | 5.5 | 1.25 | 740 | 0.17 |
| (6) | 568 | (0.531; 0.466) | 56.0 | 376 | 5.0 | 1.49 | 987 | 0.31 |
| (7) | 525 | (0.280; 0.570) | 84.0 | 163a | 7.0 | 0.87b | 365b | - |

As reported by ref. (43), the EL of [Ru(bpy)$_3$]$^{2+}$as a benchmark emitter exhibits an emission band at about 630 nm. In contrast, the maximum emission wavelengths of complexes (1-6), which contain the PTA ligand, are in the 562-614 nm range.

Our experiments are showing remarkable improvement in EL characteristic of Ru(PTA) (6) with respect to that of the Ru tetrazole derivatives in the ITO/PEDOT:PSS/PVK:PBD/Ru(PTA)/Al device configuration. Device characteristics for all Ru(PTA) complexes (1-6) are given in Table 2.Upon the application of an electric field of 5 V to device (6), an increasing in the current density is detected(Figure 6 a)), reaching a luminous efficiency of 1.4 cd/A (Figure 6 b)). On the contrary, devices based on Ru (1-5) have shown a lower current density and luminous efficiency even at higher applied voltage (*e.g.*, 7.0 V), with respect to device (6), see Figure 6 a) and Figure 6 b).Moreover, device(6) shows a significant reduction in turn-on voltage, with respect to devices based on other Ru tetrazole compounds.[41]

## The influence of counter ion on EL performance

The role of counter ion on the brightness stability of Ru-based complexes is a key factor for device characteristics. For this purpose, luminance over time of the as-produced devices is tested and reported in Figure 6c. Noteworthy, the devices based on complexes containing counter ion $BF_4^-$ show higher luminance stability, if compared with complexes without counter ion (1 and 3), which, on the contrary, show the lowest luminance stability amongst the as-prepared devices.[44] One disadvantage of complexes without counter ion, which contain NCS group, is the conversion of _-NCS in $[(L)_2Ru(NCS)_2]$ groups to –NC group in $[(L)_2Ru(NC)_2]$ upon illumination. In fact, the electronic configuration of ruthenium changes from $d^6$ to $d^5$.[45-47] Ligand additivity, the formation of L2Ru(NCS)(CN) complex, is ruled out because its potential would be expected to lie between those for the thiocyanate and cyanide complexes at +0.51 V. However, even when the applied oxidation potential for the thiocyanate complex is set at +0.27 V, only the formation of the bis cyanide complex is observed. This result implies that the sulfur elimination reaction may occur via formation of a cyclic intermediate and structural rearrangement, as the thiocyanate ligand flips from N bond to C bond after the loss of sulfur (Please see ESI, S5 for more details).[48]

The as-prepared and tested devices(*i.e.*, Ru(PTA)-based) show better performances in term of voltage, luminous efficiency of 1.49 cd A$^{-1}$at 12 V and luminance of 987cd m$^{-2}$at 12 V, if compared with the ones achieved recently with Ru(TzBA) (TzBA=4-(1H-tetrazole-5-yl)benzoic acid containing potassium (K$^+$) counter ion.[49] The latter has indeed shown luminous efficiency of 0.87 cdA$^{-1}$ and luminance of 365 cd m$^{-2}$ at 16 V. These data clearly show a twofold improvement of the emitting properties of Ru(PTA)-based together with a 25% decreases of the applied voltage. Although, some ruthenium polypyridyl complexes have shown high brightness and efficiency OLED,[50-53] the EL performance of Ru(PTA) complexes is higher than the common ruthenium polypyridyl complexes.[54-70] (Please see the ESI, S7 for details). The EL efficiency here reported is the highest recorded in all ruthenium tetrazole family tested so far (please see Table S1).The difference in turn-on voltage between device (6) and the others (1-5), see Figure6, can be explained by considering the charge transport mechanism through counter ion mobility.[71,72]

In device (6), the high current density suggests that the concentration of [Ru(PTA)(bpy)(pyTz)]BF$_4$ sites in the blend is sufficient to provide the conducting channel for charge injection and transport. The energy transfer process in ruthenium-based light emitting diode is of great interest. Here, Förster resonance energy transfer can happen through Columbic interaction (Förster) mechanism and two major conditions have to be satisfied. First, the spatial separation between the donor and acceptor when an excited donor transfers energy over to a ground-state acceptor, has to be in the 1-10 nm range.[73] Second, there should be spectral overlap between the donor emission and the acceptor absorption.[74,75] The PVK:PBD blend with emission band at 455 nm can transfer electron to Ru(PTA) complex as electron acceptor with maxima absorption band at about 450 nm. The appropriate position of HOMO and LUMO of the different layers allows the electron transfer from PVK:PBD host to Ru(PTA) complexes, confirming the energy transfer by Förster mechanism.[76] In Ru(PTA) devices, by applying bias, holes from PVK are nearest the $BF_4^-$, while electrons injects from the cathode into $t_{2g}^*$ of metal.[77] These electro-generated ions hop through the electrodes until they form excited cations.

The process can be described as follow:

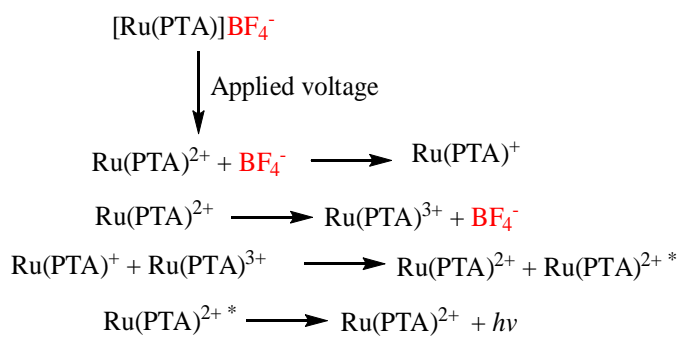

Under an applied bias, $BF_4^-$ counter ion in Ru(PTA) complexes drifts, leading to the accumulation of negative counter-ion and cationic Ru complex in proximity of holes and electrons, respectively.

As reported previously,[78] the voltage increase determines a $[Ru(PTA)]^+$ cation build-up enhancement at the cathode and a $BF_4^-$ anion build-up at the anode. However, the magnitude of this effect is strongly dependent on the molecular structure of the emitters.[79] The enhancement of injected electronic charge into the orbitals of complex is originated from high electric fields at the electrodes from the presence of ionic space charge.[80] The aforementioned processes are schematized in Figure 7. Moreover, the ion transport of $BF_4^-$ is fast enough to attain a steady state.[81] In comparison with previous studies on $ClO_4^-$ and $PF_6^-$ as counter ions,[82] the difference in turn on voltage, between the devices based on the aforementioned counter ions, decreases in order: $BF_4^- > ClO_4^- > PF_6^-$. This is probably due to the solvation shell for small counter ions such as $BF_4^-$, which is larger if compared with the ones of big ions (*e.g.*, $PF_6^-$), demonstrating the difference in the rate transport of ions in solid environment.

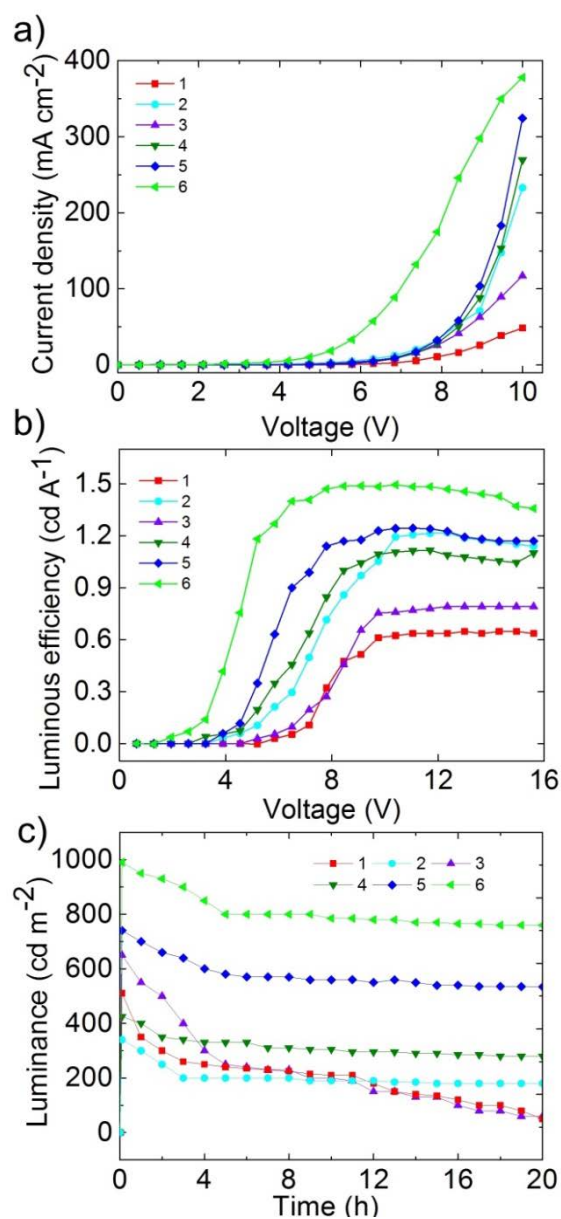

**Figure 6.** a) Current density versus applied voltage for devices (1-6). b) The maximum luminous efficiency (LE) versus applied voltages (V) characteristics of Ru-based devices (1-6). c) Luminance over time for devices(1-6).

Therefore, we can expect that the slow turn on voltage reported in [Ru(TzBA)(bpy)(pyTz)(SCN)]K, 4-(1H-tetrazole-5-yl) benzoic acid (TzBA)-based device, seeRef. (30), is due to the low mobility of the K$^+$ ions at the anode. This problem can be solved by changing the (K$^+$) counter ion with higher mobility ones, such as BF$_4^-$ in [Ru(PTA)(bpy)(pyTz)]BF$_4$ (6) (see table 2).However, the aim of this work is not devoted at the obtaining of record EL efficienciesin OLED, which can be achieved by the incorporation of LiF cathodes[83] and other efficient electron and hole transport materials.[84] On the contrary, our work is focused towards the understanding of the modifications of ancillary ligands as well as the use of counter ion in the improvement of the electron mobility, compared to neutral complexes and potassium counter ion previously reported.[30,43,49]

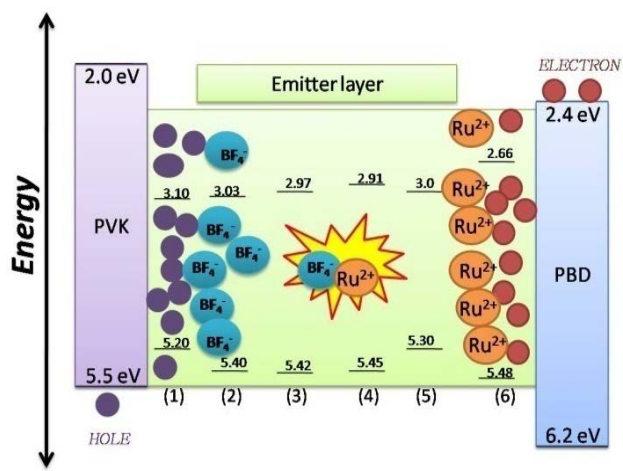

Figure 7. Schematic representation of the device energy levels and the dynamic process of the EL emission. The schematic illustration assumes conducting hole transport layer(HTL) and electron transport layer (ETL) with emitter layer, showing charge tunneling through the electrode-organic interface barrier. The negative counter ions are displaced when an external bias is applied.

## CONCLUSIONS

In conclusion, we have demonstrated that ionic Ru tetrazole complexes can be used to engineer light-emitting electrochemical cells, by using a $BF_4^-$ counter ion. Although the aim of this work was not focused to the search of the best performances and to maximize the cell efficiencies, we haveshown how the modifications of the ancillary ligands and counter ion contribute to improve the overall performance of light emitting diodes. The obtained results indicate that $BF_4^-$ functionalized Ru tetrazole complexes could boost the optoelectronic performances of next-generation light-emitting electrochemical cells.

## ACKNOWLEDGMENTS

FB was supported by the European Commission through the Graphene Flagship‐ Core 1 project (GA-696656).

# Ruthenium Tetrazole-Based Electroluminescent Device: The Key Role of Counter-Ions for the Light Emission Properties

Hashem Shahroosvand[a,*], Leyla Najafi[a,b,c], Ahmad Sousaraei[a], Ezeddin Mohajerani[d], Mohammad Janghouri[d], and Francesco Bonaccorso[b,*]

[a] Chemistry Department, University of Zanjan, Zanjan, Iran
[b] Istituto Italiano di Tecnologia, GrapheneLabs, 16163 Genova, Italy
[c] Dipartimento di Chimica e Chimica Industriale, Università degli Studi di Genova, Via Dodecaneso 31, Genoa, 16146, Italy
[d] Laser and Plasma Research Institute, Shahid Beheshti University, Tehran, Iran

## S1. Photoluminescence and electroluminescent properties of EL devices

**Table S.1.** The photoluminescence (PL) and electroluminescent (ECL) properties of investigated ruthenium tetrazole complexes. $\lambda_{max}$ is the maximum emission wavelength, $\Phi$ is quantum yield, while $\tau$ is the lifetime value.

| Ruthenium tetrazolederivates | PL | | ECL | | |
|---|---|---|---|---|---|
| | $\Phi$ | $\tau$ | $\lambda_{(max)}$ | rECL-Int. (%) | Ref. |
| $[Ru(2-TT)(tpy)(bpy)]^+$ | <0.001 | 18.64 | - | - | [1] |
| $[Ru\{2-(Me)TT\}(tpy)(bpy)]^{2+}$ | <0.001 | 22.10 | - | - | [1] |
| $[Ru\{2-(H)TT\})(tpy)(bpy)]^{2+}$ | <0.001 | 18.44 | - | - | [1] |
| $[Ru(2-TBT))(tpy)(bpy)]^+$ | <0.001 | 15.79 | - | - | [1] |
| $[Ru(Me)(2-TBT))(tpy)(bpy)]^{2+}$ | <0.001 | - | - | - | [1] |
| $[(tpy)(bpy)Ru(TDT)Ru(tpy)(bpy)]^{2+}$ | <0.001 | 19.34 | - | - | [1] |
| $[(tpy)(bpy)Ru(Me)(TDT)(Me)Ru)tpy)(bpy)]^{4+}$ | <0.001 | 22.15 | - | - | [1] |
| $[Ru(4-(Me)TBN)(tpy)(bpy)]^{2+}$ | 0.008 | - | - | - | [2] |
| $[Ru(4-TBN)(tpy)(bpy)]^+$ | 0.003 | - | - | - | [2] |
| $[(tpy)(bpy)Ru(BTB)Ru(tpy)(bpy)]^{2+}$ | 0.003 | - | 710 | 45 | [2] |
| $[(tpy)(bpy)Ru(4-TBN)Ru(tpy)(bpy)]^{2+}$ | 0.008 | - | 680 | 120 | [2] |
| $[Ru(bpy)_2(pyTz)]^{2+}$ | 0.004[a] | 220 | - | - | [3] |
| $[Ru(bpy)2(pyrTz)]^{2+}$ | 0.003[a] | 6 | - | - | [3] |
| $[Ru(bpy)_2\{4-Me-(pyTz)\}]^{2+}$ | 0.001[a] | 160(10) | - | - | [3] |
| $[Ru(bpy)_2\{4-Me(pyrTz)\}]^{2+}$ | 0.003 | 150 | - | - | [3] |
| $[Ru(bpy)(tpy)(TPh)]^+$ | 0.003 | - | 740 | 75 | [4] |
| $[Ru(bpy)(tpy)(4-pyTz)]^+$ | 0.004 | - | 730 | 15 | [4] |
| $[Ru(bpy)(tpy)(4-Me(TBN)]^+$ | 0.006 | - | 720 | 45 | [4] |
| $[(tpy)(bpy)Ru(BTB)Ru(bpy)(tpy)]^{2+}$ | 0.005 | - | 700 | 45 | [4] |
| $[Ru(bpy)_3]^{2+}$ | 0.06 | - | 630 | 100 | [4] |

The relative ECL intensities (rECL-Int.) are calculated by r ECL-Int.(%)= 100* $I_{ECL, MAX}/I_{ECL, MAX, Ru(bpy)_3^{2+}}$.

**tpy**: 2,2':6', 2''-terpyridine
**bpy**: 2,2'-bipyridyl
**2-TTH**: 2-(1-H-tetrazole-5-yl)-thiophene
**2-TBTH**: 5-bromo-2-(1-H-tetrazole-5-yl)-thiophene
**4-TBNH**: 4-(1,H-tetrazole-5-yl)-benzonitrile
**BTBH$_2$**: Bis-1,4-(1,H-tetrazole-5-yl)-benzene
**pyTz:** 2-(1,H-tetrazol-5-yl)pyridine)
**PyrTz**: pyrazinyl-tetrazolate
**bpyrTz** : 2,3-bis(1,H-tetrazol-5-yl)-pyrazine

## S2. Experimental

### S2.1 Materials and methods

All reagents including solvents and materials are purchased from Aldrich & Merck companies. 1HNMR and Infra-red spectra are recorded with a Bruker 250 MHz and Perkin-Elmer 597 spectrometers, respectively. Cyclic Voltammetry (CV) analysis is performed by potentiostat model SAMA500 (ANGHAM HOSSEYNI Com.).

For CV analysis, a conventional three-electrode having a glassy carbon as working electrode and Pt-wires as both counter and reference electrodes. All experiments are carried out on acetonitrile solvent in room temperature with 0.1 M TBATFB as supporting electrode. Ferrocene-ferrocenium (Fc/Fc$^+$) couple is used as an internal standard for the potentials reported for each experiment at a scan rate of 100 mV/s. The oxidation ($E_{ox}$) and reduction ($E_{red}$) potentials are exploited for the determination of both the highest occupied molecular orbital (HOMO) and lowest unoccupied molecular orbital (LUMO) energy levels by using the following equations: $E_{HOMO} = -(E_{ox} + 4.8)$ eV and $E_{LUMO} = -(E_{red} + 4.8)$ eV. The value of -4.8 eV is the internal standard ferrocene with respect to vacuum.[5,6]

### S 2. 2. Synthesis of ligand and complexes

*Caution: Azide and tetrazolate compounds are potentially explosive. The reactions described here were carriedoutusing a few grams of both compounds and we did not encounter problems. However, caution should be taken when handling or heating this typology of chemical agents.*

2-pyridine (1H-tetrazole-5-yl) (pyTz) is prepared according to the literature method.[7]

**PTA ligand.** PTA ligand is prepared from the reaction between pyTz(0.05 mmol, 0.0043g) and chloroacetic acid (0.05 mmol, 0.1g) in 150 ml ethanol solution under reflux for 24 hours. The resulting reaction mixture is allowed to cool and the crystals in ethanol solution are then filtered and washed carefully with methanol and acetone solutions. Anal. Calc. for PTA , ($C_8 H_7 N_5 O_2$) : C, 55.492; H, 4.07; N, 40.443. Found: C, 55.50; H, 4.09; N, 40.15%. ESI-MS: m/z, 172.20, [M-H]$^-$. **PTE ligand.** PTE ligand is prepared by esterification of PTA through the dissolution of PTA (1mmol, 0.206 g) in the mixture of H$_2$SO$_4$ (1ml, 3N) and 150 ml of methanol. For purification, the crystals are filtered and washed with water and acetone, several times. Anal. Calc. for PTE, (C$_9$ H$_9$ N$_5$O$_2$) : C,49.313; H, 4.144; N, 31.98. Found: C, 49.320; H, 4.418; N, 40.450%. ESI-MS: m/z, 218.21, [M-H]$^-$.

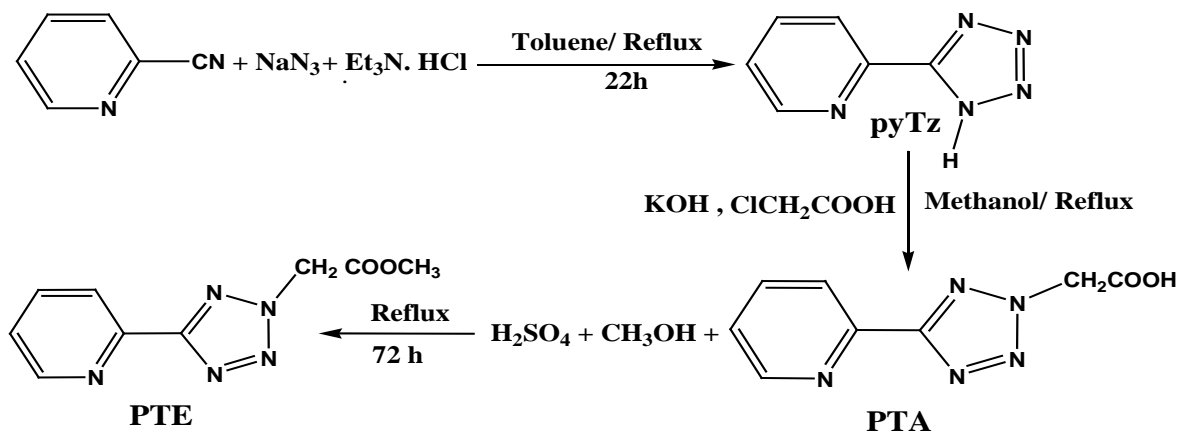

**Scheme S 2.1:** Synthesis procedure of pyTz, PTA and PTE ligands.

Scheme S2.1 summarizes the synthesis procedure of (pyTz), (PTA) and (PTE), while Scheme S2.2 shows the synthesis pathway of the heteroleptic complexes.

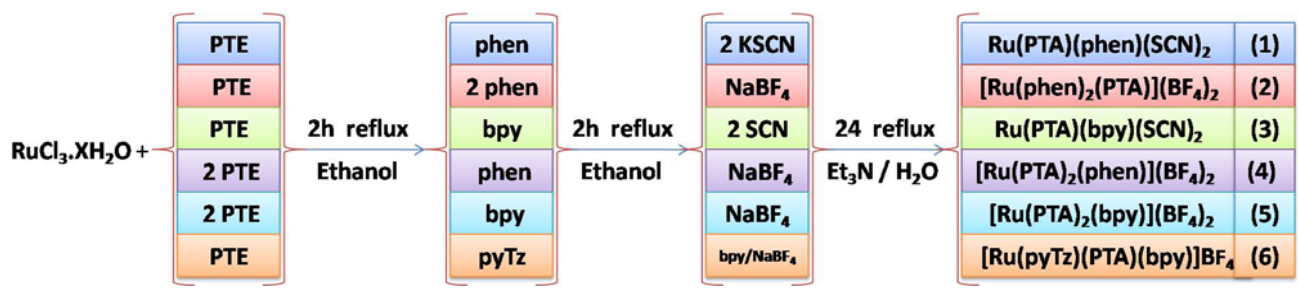

**Scheme S2.2**: Synthesis pathways for Ru(PTA) complexes (1-6)

Complex Ru(PTA)(phen)(SCN)$_2$ (1). For the synthesis of complex (1), RuCl$_3$.H$_2$O (0.1 g, 0.16 mmol) is dissolved in ethanol (50 ml) and to this solution 1eq of PTE ligand is added. The reaction mixture is subsequently refluxed under nitrogen constant stirringfor 2 h. Afterward, 1 eq of phen as anicillary ligand is added to this reaction mixture and refluxed for 2h. Finally, 2 eq potassium thiocyanate (KSCN) is added to the mixture. The solid products are washed with distilled water and diethyl ether for three times. The conversion of an ester group of complexes to acidic group is carried out by adding 20 ml of triethylamine (TEA) in 10 ml of water to complex (6), and kept under reflux for 24 hours. After the cooling of mixture, the filtering of the obtained dispersion is carried out by centrifugation at 5000 rev/min for 30 minutes. The washing of solids with water, acetone and ether is performed to remove any un-reacted reagents. The purification of products is also carried out by using Sephadex LH20 as the column support and acetonitrile/methanol (2:1, v/v) as the eluent. Anal. Calc. for (1) , ($C_{22} H_{15} N_9 O_2 S_2 Ru$) : C,43.813; H, 2.615; N, 20.808. Found: C, 43.829; H, 2. 609; N, 20. 819%. 1HNMR, 8.7(d,a1*), 8.6 (t,a2), 8,48(d,b1), 8.3(d,a1*), 8.17(d,a4,a4*), 8.02(d,a3,a3*), 7.979t,a2*), 7.8(q,b2,b3), 7.67(d,b4).ESI-MS: m/z, 601.6, [M-H]$^-$.

Complex Ru(PTA)(phen)$_2$(BF$_4$)$_2$ (2). The procedure of synthesis of complex (2) is similar to complex (1) except for the use of 2 eq of phen as ancillary ligand and the replacement of KSCNwith BF$_4$. Anal. Calc. for (2) , ($C_{32} H_{23} N_9 O_2 RuB_2F_8$) : C, 45.742; H, 2.763; N,15,.010. Found: C, 45.749; H, 2.770; N, 15..00%.H, 2.743; N, 21.627%. 1HNMR, 8.8(d,a1*), 8.64(d,b1), 8.62(d,a1**), 8.47(d,a1), 8.42(d,a3), 8.27(q,a2), 8.24(d,a4*), 8.2(q,b2,b3), 8.1(d,a3), 8.05(d,a4), 7.8(q,a2**), 7.57(d,b4). ESI-MS: m/z, 665.7, [M-H -2BF$_4$]$^+$.

Complex[ Ru(PTA)(bpy)$_2$(SCN)$_2$]  (3). The complex (3) is synthesized according to procedure of complex (1) with the only difference relying in the replacement of 1eq of phenwith 1 eqof bpy as ancillary ligand. Anal. Calc. for (3) , ($C_{20} H_{15} N_9 O_2 S_2 Ru$) : C, 41.531; H, 2.790; N, 21.604. Found: C, 41.518; 1HNMR, 8.6 (d,a1), 8.63(d,b1), 8.5(d.a1*), 8.32(t,a,b2), 8.2(d,a2*), 8.1(q,a4,a4*), 8.05(q,a3,a3*), 7.8 (d,b4), 7.46 (t,b3)ESI-MS: m/z, 577.6 , [M-H]$^-$.

Complex [ Ru(PTA)2(phen)(BF4)$_2$]  (4). The synthesis procedure of complex (4) is similar to the one carried out for (1), except for the use of 2eq of PTA and the replacement of 2 eq of KSCN with 2eq of BF$_4$. Anal. Calc. for (4) , ($C_{28} H_{22} N_{12} O_4 RuB_2F_8$) : C, 38.873; H, 2.565; N, 19.433. Found: C, 38.880; H, 2.572; N, 19.442%. 1HNMR, 8.79 (d,a1), 8.77(d,b1), 8.23(t,a2), 8.21(t,b2), 8.09 (d,a4), 8.05(d,a3), 8.0 (t,b3), 7.61(d,b4)ESI-MS: m/z, 691.6, [M-H-2BF$_4^-$]$^+$.

Complex [ Ru(PTA)2(bpy)(BF4)$_2$]  (5). The synthesis of complex (5) is similar to the one of complex (4) except phen  was replaced by bpy as ancillary ligand. Anal. Calc. for (5) , ($C_{26} H_{22} N_{12} O_4 RuB_2F_8$) : C, 37.121; H,  2.644; N, 19.982. Found: C, 37.128; H, 2.650; N, 19.975%. 1HNMR, 8.82(d,a1), 8.72(d,b1), 8.55(t,a2), 8.42(t,b2), 8.32(d,a4), 8.25(d,a3), 7.9(d,b4), 7.65(t,b3). ESI-MS: m/z,665.6, [M-H-2BF$_4^-$]$^+$.

Complex [ Ru(pyTz)2(PTA(bpy)][(BF4)]  (6), the synthesis of complex (6) is similar to complex (5) except 1 eq of all four components including  pyTz, PTA, bpy and BF$_4$are used. Anal. Calc. for (6), ($C_{24} H_{19} N_{12} O_2 RuB_2F_8$): C, 41.461; H, 2.750; N, 24.1714. Found: C,41.428 ; H, 2.755; N, 24.179%. 1HNMR, 8.77(d,a1),

8.74(d,b1), 8.27(t,a2), 8.25(t,b2), 8.16(t,c2), 8.14(t,a3), 8.03(d,c1), 8.02(d,a4), 7.989d,a4*), 7.88(t,a3*), 7.58(t,a2*), 7.56(d,b4), 7.55(d,c4), 7.54(d,a1*), 7.36( t,c3,b3). ESI-MS: m/z 595.5, [M-BF$_4$]$^+$.

## S3. Characterization

### S3.1. Absorption and photoluminescence studies

The absorption spectra of the Ru(PTA) complexes, measured in acetonitrile solution at roomtemperature, show intense bands in the ultraviolet region, below 300nm, see Figure 3 of the main text. These bands are attributed to the spin allowed n, $\pi \to \pi^*$ transitions of the ligands,[8] which are analogous to the one of tetrazole-based complexes.[9] In addition, as typically shown by Ru(II)-polypyridine chromophores,[10] each spectrum shows the metal-to-ligand charge-transfer (MLCT) transitions in the ~400-600nm absorption range.

The PL spectraare recorded by Avantes spectrometer Ava-Spec128 with a 405nm excitation wavelength. The excitation spectrum of complex (1) in acetonitrile solution at RT is shown in Figure S1.

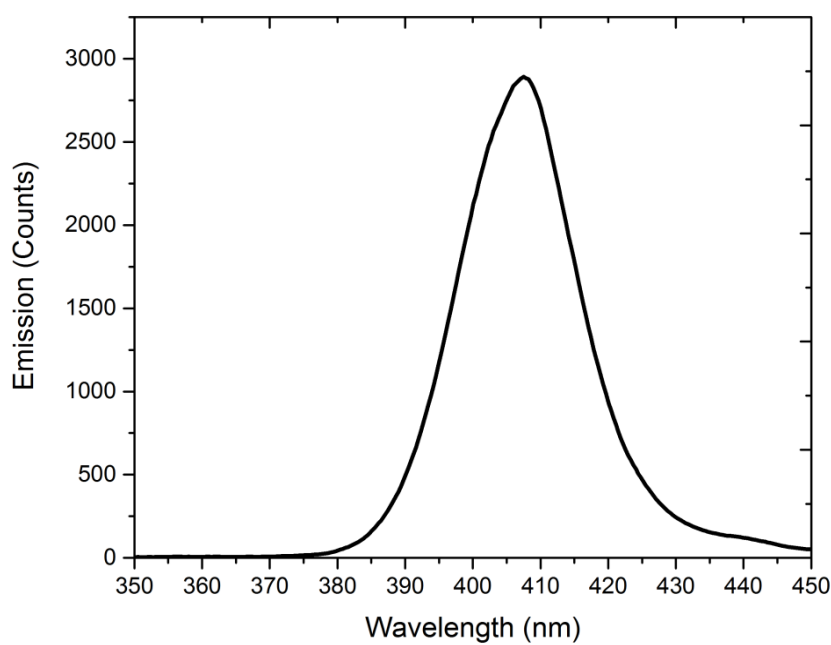

**Figure S 1.** The excitation spectrum of complex (1) in acetonitrile solution at RT. Excitation wavelength is 405 nm.

### S3.2. $^1$HNMR spectroscopy

$^1$H-NMR spectroscopy is a powerful tool to investigate the contribution of ligand in the synthesized complex, see Figure S.2.1-S.2-6. The major ligand to be investigated by $^1$H-NMR is the PTA ligand. Through the coordination of PTA to ruthenium metal, some up-field shifts are expected. In this context, differentfeatures contribute to the coordination-induced shift values (CIS = δ complex- δ ligand) in such ruthenium (II) complexes.[11-13]

Chelation-imposed conformational changes, ligand-to-metal donation, coordinative disruption of inter-ring conjugation, metal-to-ligand back donation and through-space ring-current anisotropy effects have all been investigated by $^1$HNMR spectroscopy to explain CIS values of heteroleptic polypyridine ruthenium(II) complexes. Typically, the $^1$H-NMR spectrum of (6) shows 16 non-equivalent protons (2-substituted pyridine rings) due to the un-symmetrical nature of the pyTz ligand.[1,14,15] Although the position of signals for PTA ligand could be not assigned undoubtedly, the significant insights can be achieved by discriminating the ancillary ligand through the replacement of bpy with phen or pyTz ligands. The typical $^1$H NMR of Ru(PTA)(6) complex is shown in Figure S2. 6. For complex (6), $Ha_1$ and $Hb_1$ (Carbon's Hydrogens are the nearest distance to nitrogen of pyridine ring) indicated the largest down field CIS due to the protons that are close to the second pyridine units,[16,17] whereas the protons that are in the vicinity of non-bipyridine ligands locate at the high field proton resonances.[2,8,18-21] The origin of the de-shielding of the protons can be attributed to an induced magnetic field created by the ring current circulation on py's aromatic rings. This de-shielding is only significant at short distance and therefore only affects protons in close proximity to bipyridine.[22,23] Therefore, $Ha_1$, $Ha_1^*$ protons are not equivalent. This un-equivalency can be attributed to the $Ha_1$ and $Ha_1^*$ affected by N atoms of bpy and N atoms of tetrazole moieties, respectively. The same is valid also for the $Ha_2$ and $Ha_2^*$, which are also not equivalent. $Ha_4$ and $Ha_4^*$ protons are affected by the meta effect of the N coordinated to the metal. Therefore, $Ha_4$ and $Ha_4^*$ protons appear de-shielded and their coupling with $Ha_2$ and $Ha_2^*$ is strongly affected. The other protons (remote hydrogens from N atoms), experience smaller upfield shifts due to a combination of factors. Therefore, the observed pattern is in agreement with the vicinity of one pyridine ring of bpy to pyridine ring of PTA and another pyridine ring of bpy to tetrazole moiety of pyTz ligand.[4,24,25] A. J. Downard et al. reported that negative CIS values may be a consequence of deprotonation of the tetrazole and carboxylic acid.[26] In fact, electron density of negatively charged tetrazole transfers to pyridine ring, which is a n-deficient moiety.[26] The signal integration for all complexes reveals the incorporation of ligands as predicted. The sharp resonance indicates the diamagnetic behavior of Ru(II) complexes with $t_{2g}^6$ configuration.[27-29]

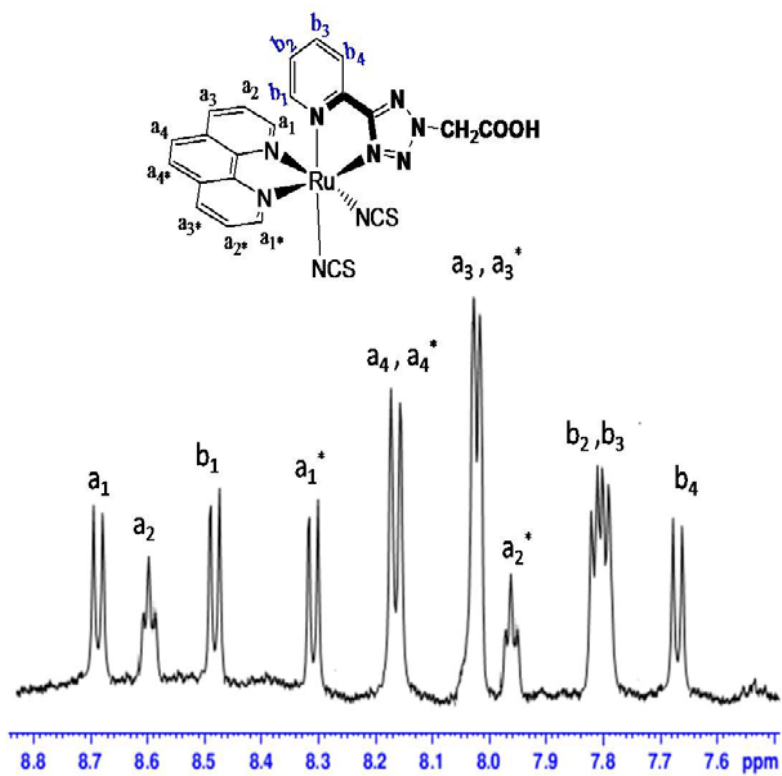

**Figure S2. 1.** Aromatic region of $^1$H NMR spectra (400MHz) of complex (1) in DMSO-$d_6$ at RT.

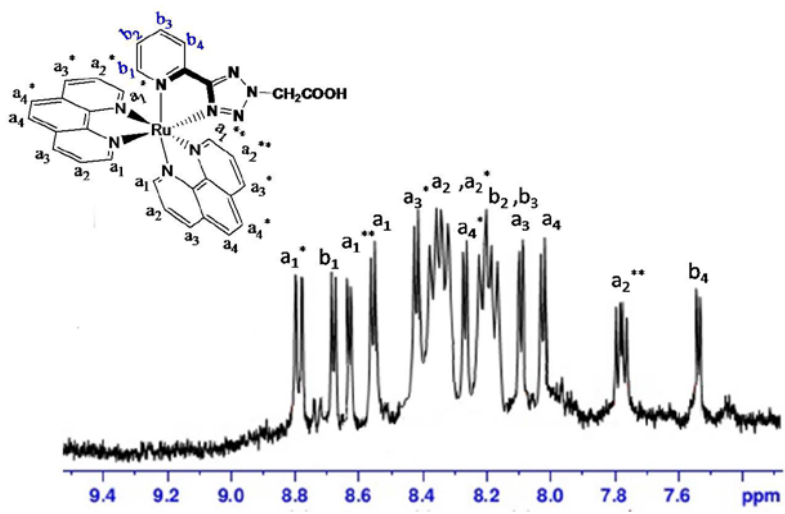

**Figure S2. 2.** Aromatic region of $^1$H NMR spectra (400MHz) of complex (2) in DMSO-$d_6$ at RT.

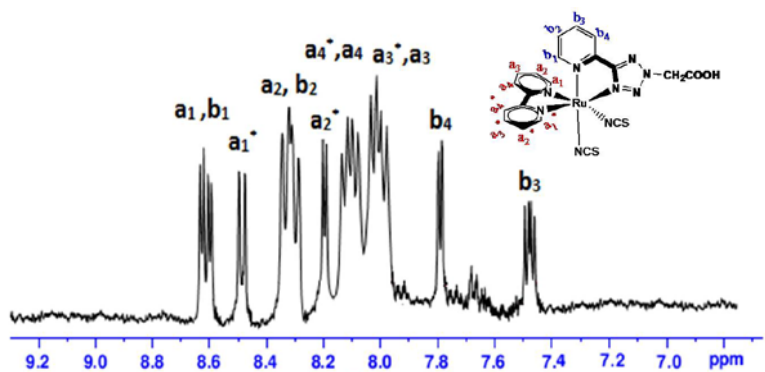

**Figure S2. 3.** Aromatic region of $^1$H NMR spectra (400MHz) of complex (3) in DMSO-d$_6$ at RT.

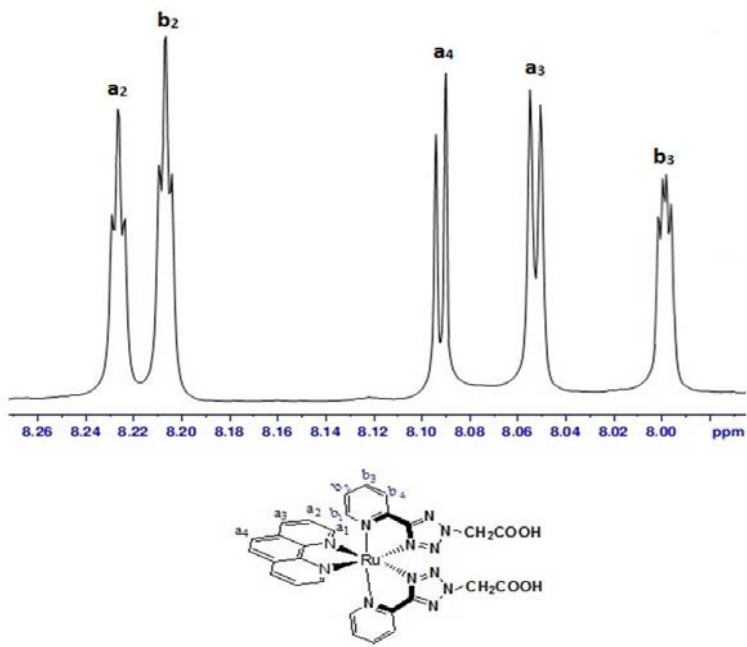

**Figure S2. 4.** Aromatic region of $^1$H NMR spectra (400MHz) of complex (4) in DMSO-d$_6$ at RT.

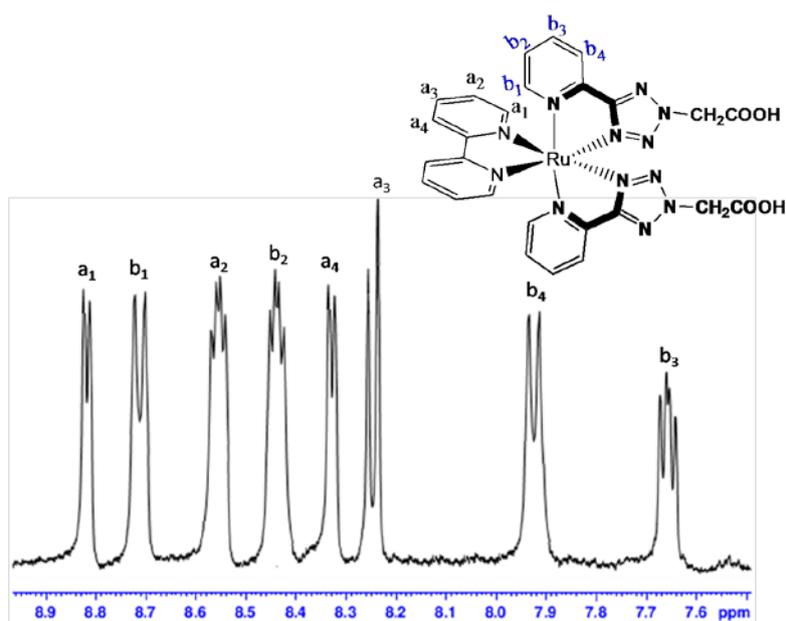

**Figure S2. 5.** Aromatic region of $^1$H NMR spectra (400MHz) of complex (5) in DMSO-d$_6$ at RT.

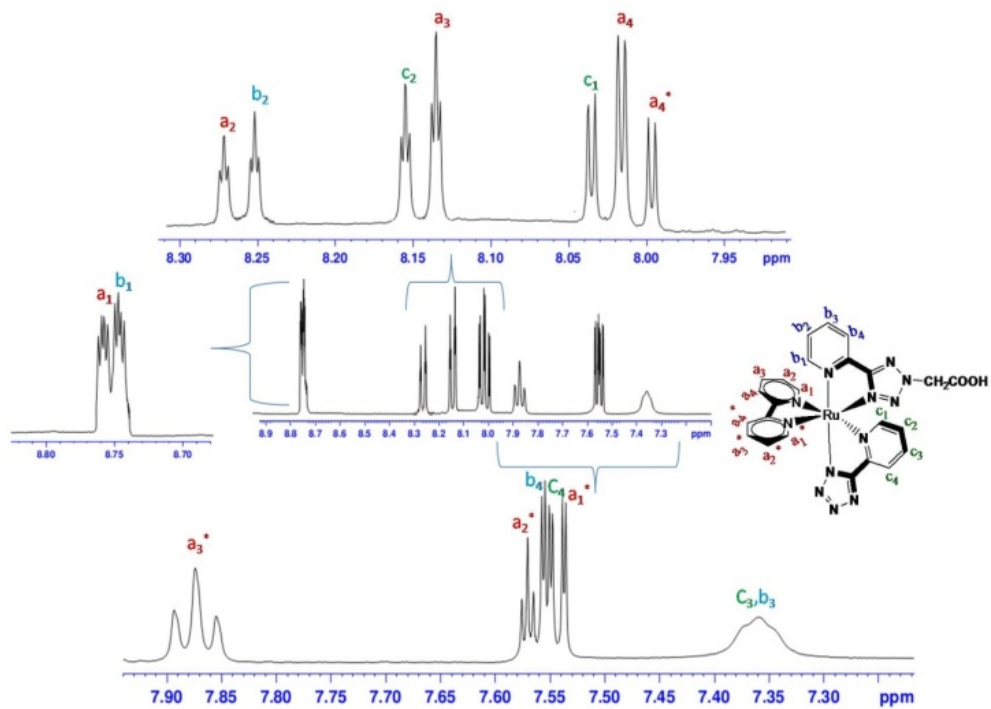

**Figure S2. 6.** Aromatic region of $^1$H NMR spectra (400MHz) of complex (6) in DMSO-d$_6$ at RT.

## S3.3. FT IR spectroscopy

The comparison between simulated IR spectra and experimental results is shown in Figure S3.

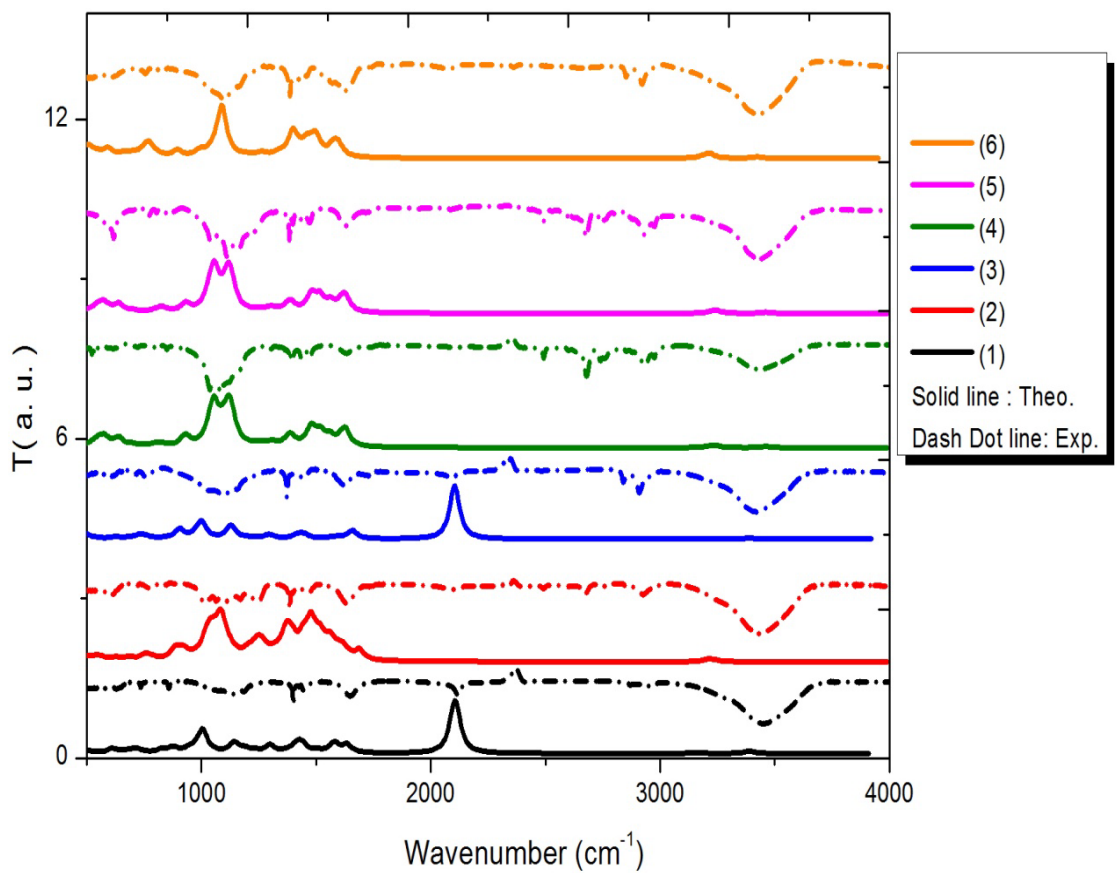

**Figure S3.** The simulated (solid lines) and experimental (dash dotted lines) FT-IR spectra of complexes (1-6).

There is a good agreement between the obtained results of computational calculation and the experimental analysis. The IR spectrum of PTA shows aseries of peaks in the 700-1600 cm$^{-1}$ region, which are characteristic of aromatic rings.[30-35] A comparison with the IR spectrum of raw materials PTA, shows the absence of a strong peak at 2275 and 2280cm$^{-1}$ for 4-cyano pyridine, indicating the completehydrolyzation of the cyano group to tetrazole ring.[7, 36-39] The overlapping between the vibrational signals of various types of C-O bonding, aromatic rings and carboxylate groups in the 1000-1800 cm$^{-1}$ regiongives origin to a crowded region.[40-43] The IR spectra of complexes are different from that of the ligands. Theshift of the tetrazole ring ν(C–H) stretching band from 3160-3050cm$^{-1}$ in the free PTA and pyTzligand to the lower frequencies in the spectra of their complexes (2950cm$^{-1}$) can be attributed to the coordination of the tetrazoleto metal center.[1-4] The broad peak at~3400cm$^{-1}$ in the experimental spectra can be attributed to OH stretching of adsorbed moisture into KBr disc, which are not present in theoretical IR spectra.[44-46]

## S4. DFT calculations

Density Functional Theory (DFT) using the Gaussian 03 (G03) is used for the calculation of the electronic structure, simulated FT-IR, and optimized structures. The functional with LANL2DZ basis set is performed. The symmetry of C1 or C2isused for the optimization of the structure to show the local minima on the potential energy surface.

Figure S4.1 shows the optimized ruthenium complexes. Initial calculations are carried out on the synthesized novel ruthenium complexes to determine the electron transfer mechanism. Figure S4.2 also shows the electron density contours for the HOMO and LUMO molecular orbitals of complexes (1-6) and HOMO-LUMO levels estimated based on cyclic voltammetry data, see Figure S5.1-S5.6. The determination of the relative positions of the LUMO levels for the donor and acceptor moieties isfundamental for the effectiveness of charge transport properties.[47] In fact, in complex (6) the presence of pyTz as ancillary ligand leads to increased HOMO–LUMO energy levels gap. Contrary, the presence of both phen and thiocyanate (-SCN) ancillary ligands decreases the HOMO–LUMO gap, which for complex (1) is estimated to be 2.10 eV. Thisleads to instability of HOMO–LUMO energy levels. In our calculations, complexes (2 and 4-6) have their HOMO confined in the donor moiety and their LUMO localized in the acceptor moiety. The HOMO of complexes 1 and 3 is localized in the donor moiety isothiocyanate(-NCS) ligand while, on the contrary, the LUMO is mostly confinedin the acceptor moiety.

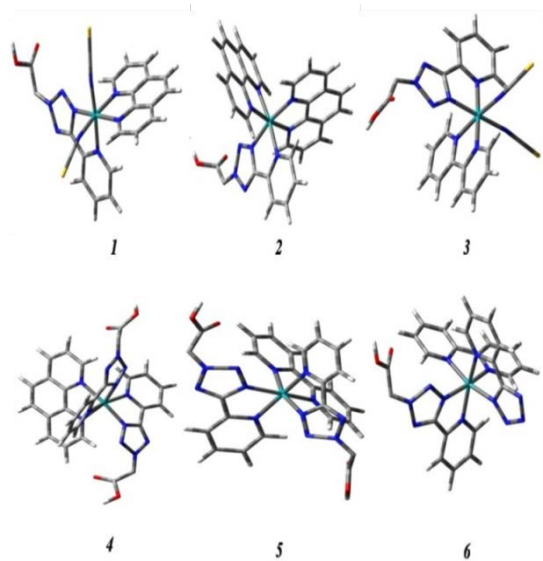

**Figure S4. 1**. Optimized structures of synthesized ruthenium complexes by using DFT based on LANL2DZ basis set.

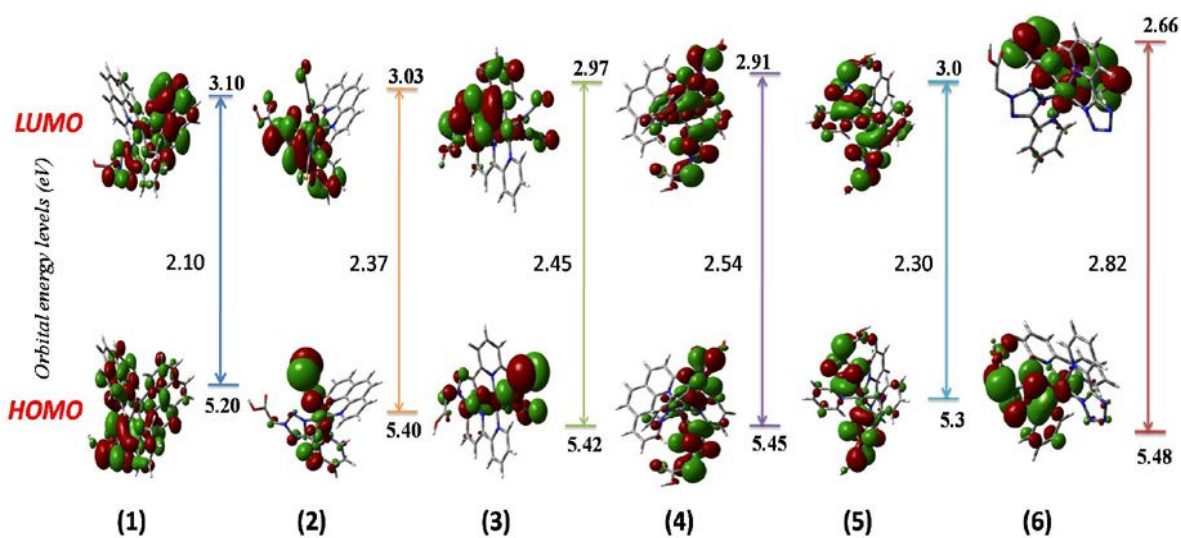

**Figure S4.2** Electron density contours for the HOMO and LUMO molecular orbitals of complexes (1-6) and HOMO-LUMO levels estimated based on CV data.

## S5. Electrochemical studies

In summary, there are two major proposed mechanisms for electron transfer and intermediate products in [L$_2$Ru(SCN)$_2$] complexes: (1) via solvent exchange, (2) via dimerization of SCN.

(1) *Solventexchangemechanism*; (i.e., radical NCS·substituted by a solvent molecule).[48] The instability of [Ru-(L)$_2$(NCS)$_2$]$^+$, which results on either intra molecular electron transfer from -NCS to Ru(III) or oxidation of ligand

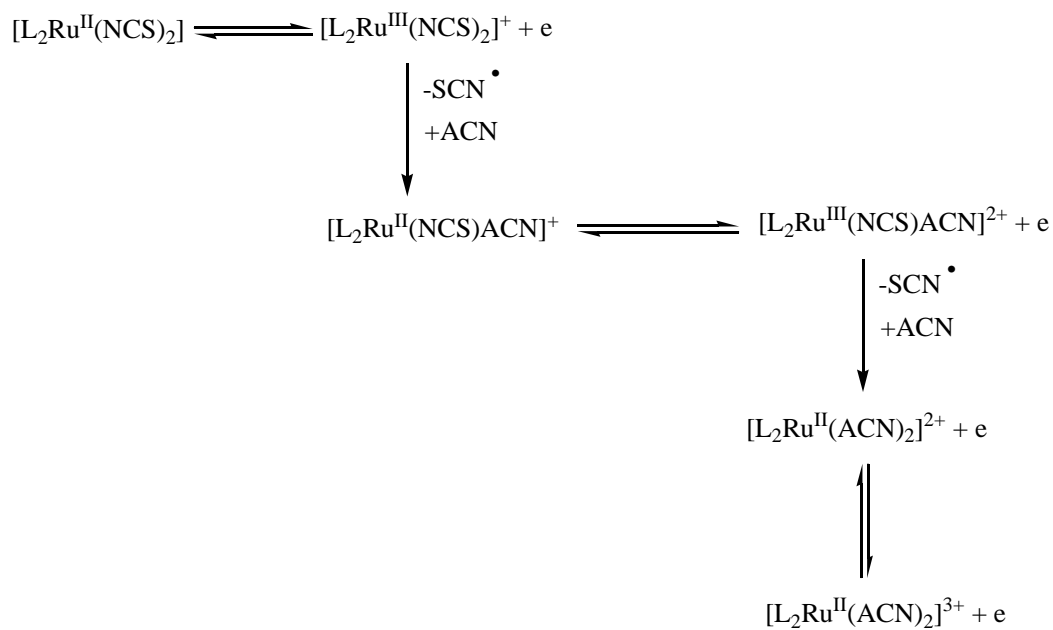

An irreversible oxidation process occurs at NCS group rather than ruthenium center. This possibility is supported by the fact that the HOMO of the complex is largely localized onto the NCS ligands.[49,50]

*(2) Dimerization of SCN;* An alternative mechanism, with respect to the solvent exchange one, stems from the work of Kohle and co-workers,[51] which have reported a slow conversion of [(L)$_2$Ru(NCS)$_2$] to the cyanide analogue upon illumination.

In fact, photoexcitation process consists of a metal to ligand charge transfer (MLCT) which causes the electron configuration from d$^6$ to d$^5$, [52-54]

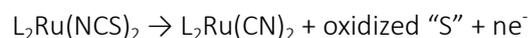

L$_2$Ru(NCS)$_2$ → L$_2$Ru(CN)$_2$ + oxidized "S" + ne$^-$

The large value of electrons transfer during the bulk electrolysis (6 ± 2 electrons per molecule) is attributed to the oxidation state of sulfur atom, which exists in the oxidation state -2 in thiocyanate[55-57] and can be oxidized up to +3, depending on the reaction pathway.[58-60]

Another possible reaction is the formation of the mixed thiocyanate-cyanide complex, L$_2$Ru(NCS)(CN).[59,60] The potential for this complex is about +0.51V which is the value between the potential of thiocyanate and cyanide complexes. However, when the potential of +0.27 is applied, only bis-cyanide, L2 Ru(NC)$_2$, complex forms. This finding indicates that the formation of a cyclic intermediate and some form of structural rearrangements. Therefore, the N-C group of the NCS group bonds to ruthenium through -C or –N atoms. Moreover, crystallographic data[61] demonstrated that two NCS groups in close proximity can lose their linear structure and the terminal sulfur could interact with other NCS group. This phenomenon could be explained considering that the NCS ligand is thermodynamically capable to reduce Ru(III) and free NCS could be oxidized at lower potential than L$_2$Ru(NCS)$_2$ complex.[62]

Overall, we have briefly given a short discussion about the dimerization of SCN to establish the irreversible oxidation process of ruthenium center.

For what concerns the redox properties of the Ru metal center, it is well-known that the two oxidation processes are possible: the first oxidation peak in ruthenium polypyridyl complexes is assigned to conversion of Ru(II) to Ru(III),[63,64] while the second one could be attributed either to a Ru (III) to Ru(IV) couple or to a ligand-based oxidative decomposition. According to the separation between the two oxidation processes, the oxidation of ligand can be assigned to the second oxidation peak.

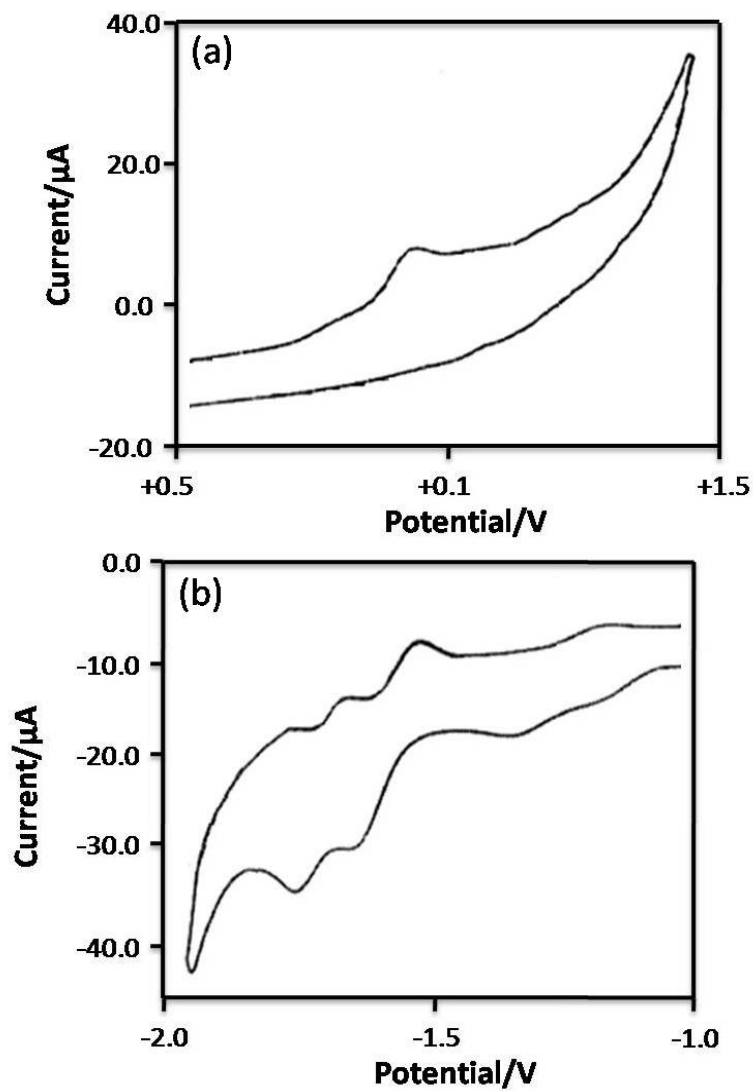

**Figure S5.1.** Cyclic voltammogram of complex (1) in $CH_3CN$ at RT under argon. The solutionconcentrationis $1.4\times10^{-3}$ M, with 0.1 M TBATFB electrolyte. (a) Ox/Red wave of Ru(II/III), (b) Ox/Red waves of ligands.

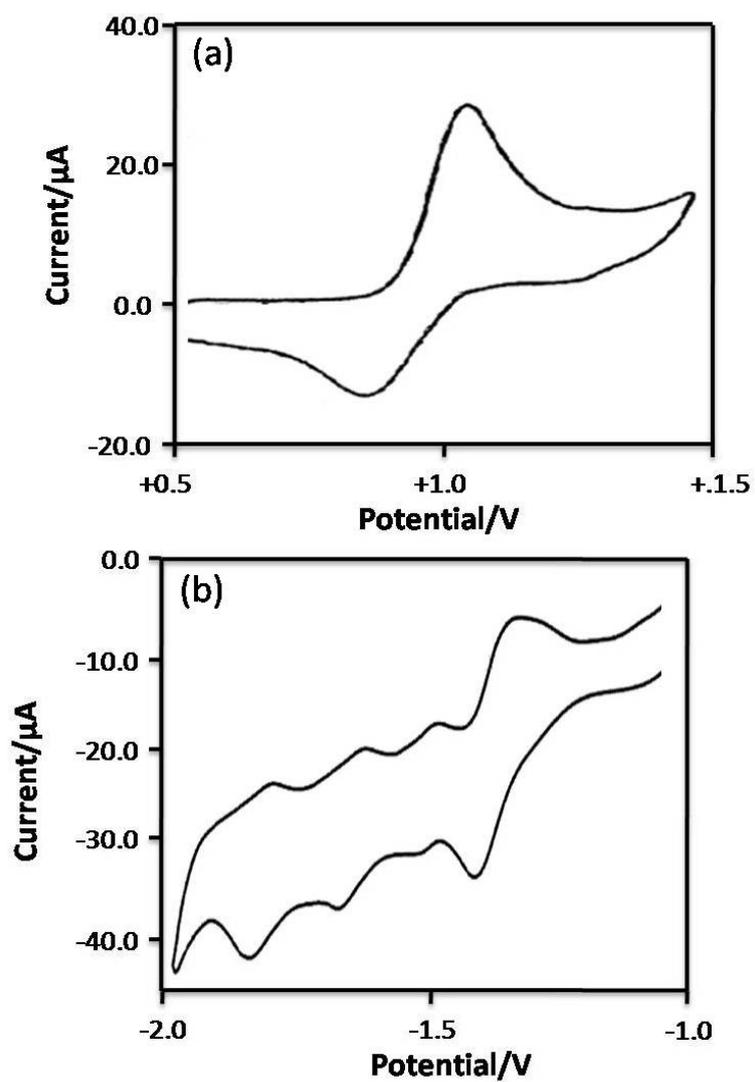

Figure S5.2. Cyclic voltammogram of complex (2) in $CH_3CN$ at RT under argon. The solutionconcentrationis 1.4×10$^{-3}$ M, with 0.1 M TBATFB electrolyte.(a) Ox/Red wave of Ru(II/III), (b) Ox/Red waves of ligands.

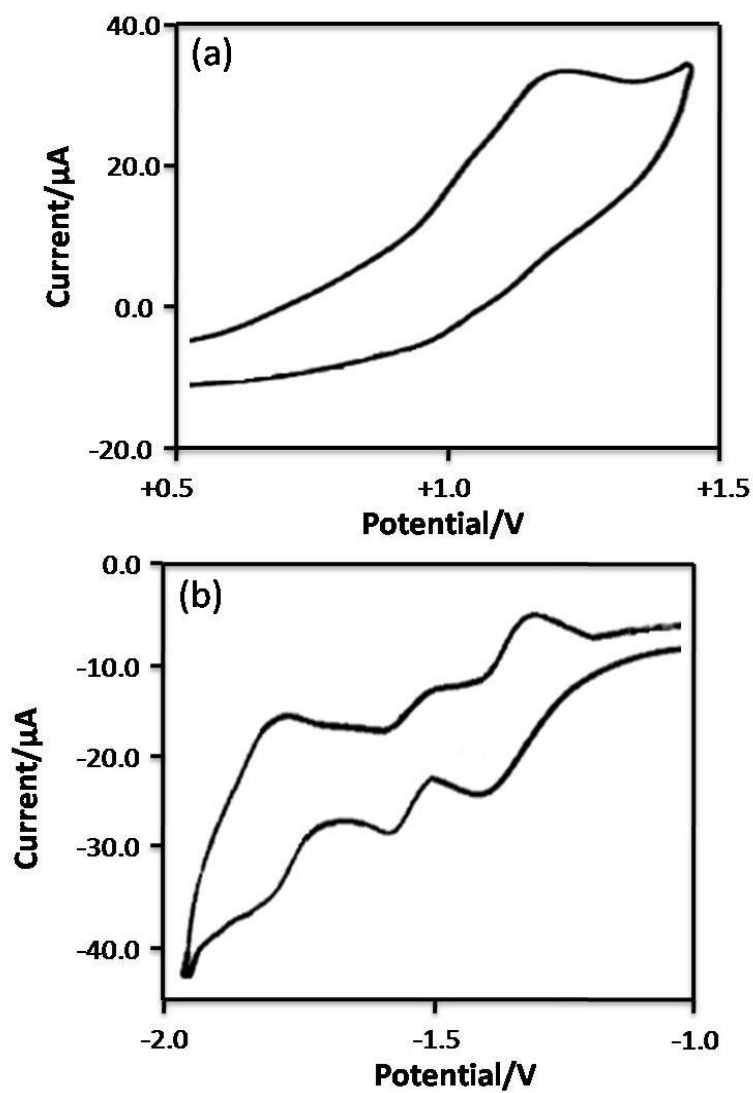

**Figure S5.3.** Cyclic voltammogram of complex (3) in $CH_3CN$ at RT under argon. The solutionconcentrationis $1.4\times10^{-3}$ M, with 0.1 M TBATFB electrolyte.(a) Ox/Red wave of Ru(II/III), (b) Ox/Red waves of ligands.

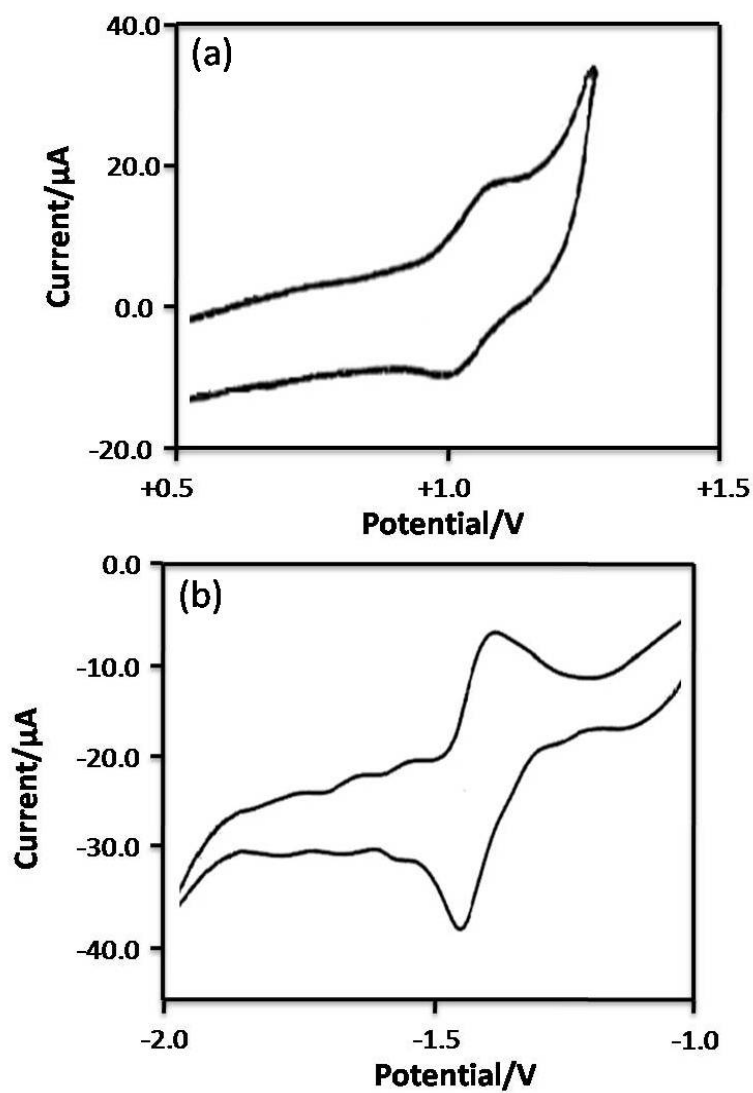

Figure S5.4. Cyclic voltammogram of complex (4) in $CH_3CN$ at RT under argon. The solution concentration is $1.4 \times 10^{-3}$ M, with 0.1 M TBATFB electrolyte. (a) Ox/Red wave of Ru(II/III), (b) Ox/Red waves of ligands.

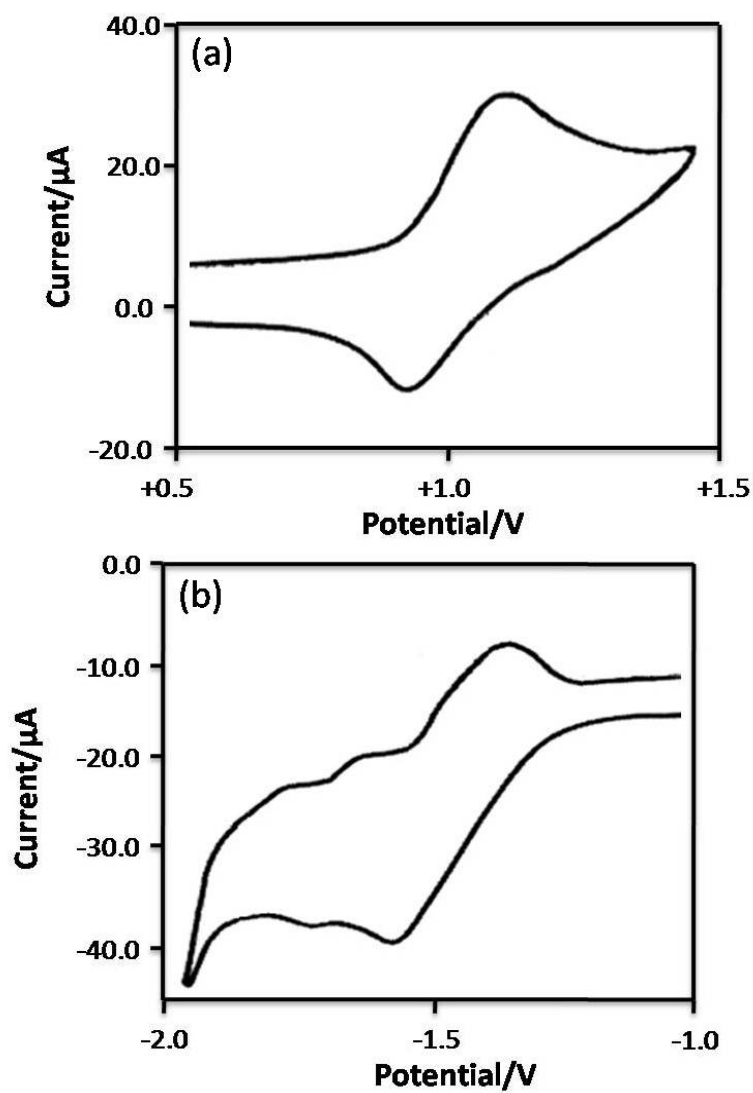

**Figure S5.5.** Cyclic voltammogram of complex (5) in $CH_3CN$ at RT under argon. The solutionconcentrationis $1.4\times10^{-3}$ M, with 0.1 M TBATFB electrolyte.(a) Ox/Red wave of Ru(II/III), (b) Ox/Red waves of ligands.

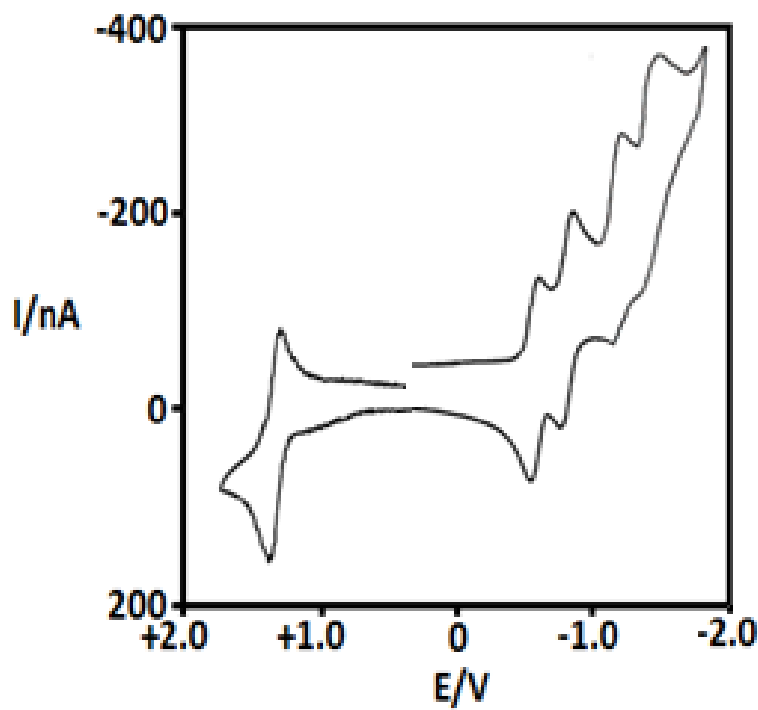

**Figure S5.6.** Cyclic voltammogram of complex 6 in $CH_3CN$ at RT under argon. The solutionconcentrationis $1.4\times10^{-3}$ M, with 0.1 M TBATFB electrolyte.

## S6. Electroluminescent devices preparation

The fabricated devices have the following structure:

Indiumtinoxide (ITO)/ (poly(3, 4-ethylenedi-oxythiophene):poly(styrenesulfonate) PEDOT:PSS(55nm)/Poly(9-vinylcarbazole) (PVK)(60nm)/rutheniumcomplex (45nm)/polybutadiene(PBD)(30nm)/aluminum(Al) (130nm), asshown in Figure S6.

Glass substrates, coated with ITO, are used as the conducting anode. The PEDOT:PSS as hole injection transporting layer is deposited onto ITO glass by spin coating method and held in oven at 120 $^0C$ for 2 hours. The spin coating conditionsare: speed of 2000 rpm for an interval time of 1 minute. PVK as a hole-transporting and PBD as an electron-transporting material are doped with ruthenium compounds. Aluminum as a metallic cathode is deposited on the complexes at $8\times10^{-5}$ bar by thermal evaporation.

The concentration of theblend of PVK, PBDRu(PTA), PTA=2-(5-(pyridin-2-yl)-2H-tetrazole-2-yl) acetic is optimized with weight ratio of 100,40, x (x=6 to 14). Each component is separately dissolved in 10 mL of acetonitrile, and then spin coated and baked at 80 °C for 1 hour.

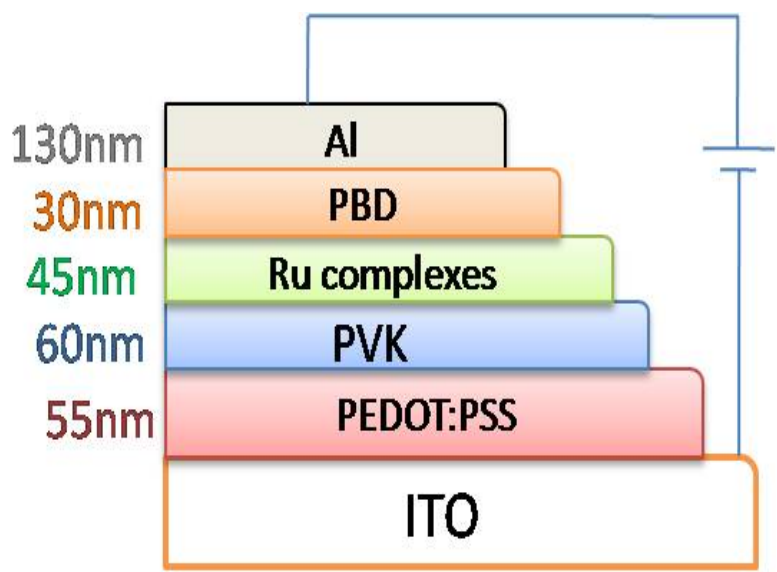

**Figure S6**. Schematic of a Ru-based LED.

The thickness of the polymeric thin film is determined by a Dektak 8000. The electroluminescence (EL) spectra are measured with an ocean optic USB2000, while aKeithley 2400 source meter is used for the measurement of the electrical properties of the as-prepared devices.

## S7. Summary of OLED characteristics of ruthenium polypyridyl complexes

In Figure S7 are reported the chemical structures of ruthenium polypyridyl complexes used in different OLED device configurations, reported in Table S.2.

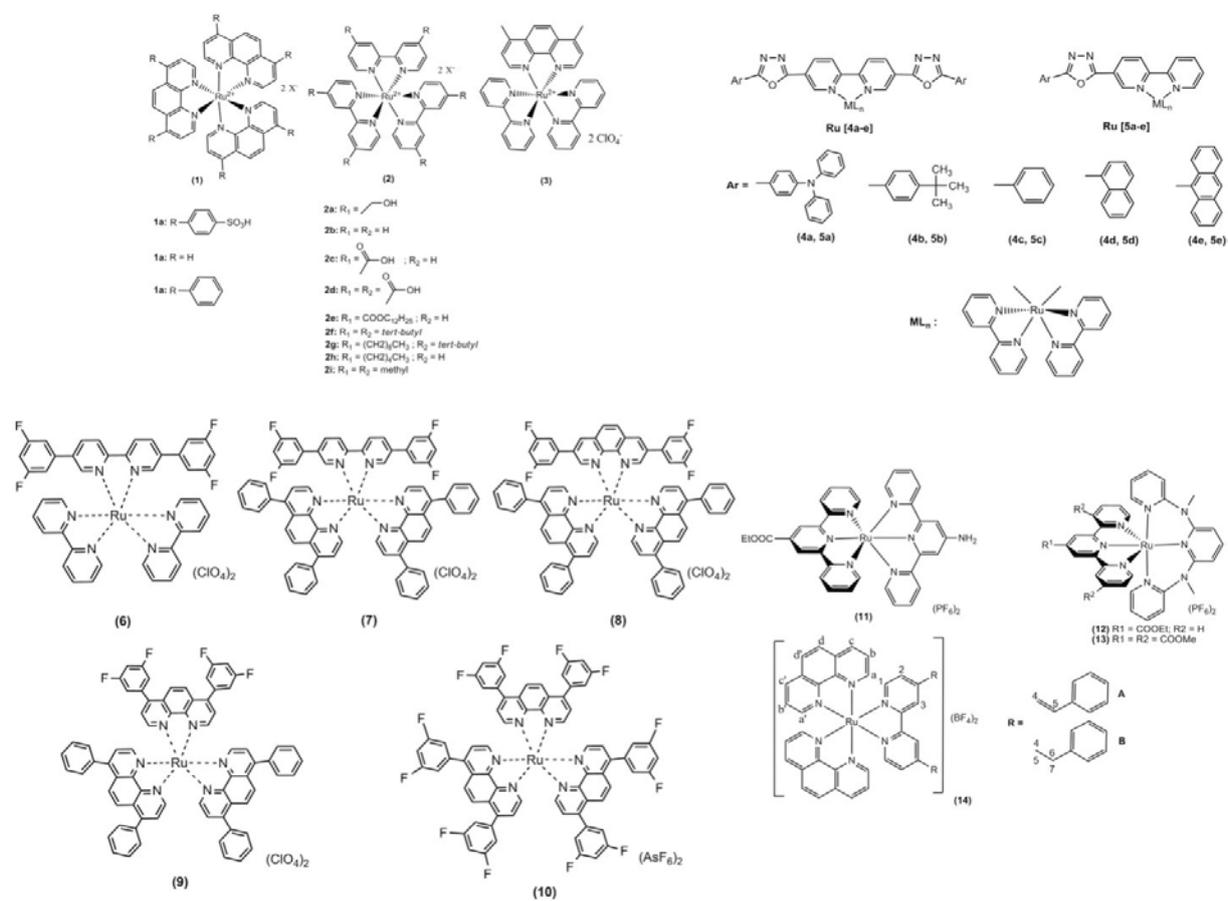

**Figure S7**. Chemical structures of ruthenium polypyridyl complexes used in different OLED device configurations. The numbers and the letters in parenthesis identify the complex used in different devices configuration reported in table S2.

Table S.2. Summary of the optoelectronic performances of OLED based on ruthenium polypyridyl complexes.

| Complex | Device Configuration | $\lambda_{max}$ abs (nm) | $\lambda_{max}$ EL (nm) | Turn-on voltage (V) | Maximum Luminance (cd m$^{-2}$) | External quantum efficiency (%) | Ref. |
|---|---|---|---|---|---|---|---|
| 1a | ITO/(PPV/SPS)$_5$/(PPV/PMA)$_{20}$/1a/Al | 470 | 628 | 2.5-3.5 | 110 | 0.05 | 65 |
| 1a | ITO/ PEO:1a:LiCF$_3$SO$_3$/Al | 470 | 628 | 2.5-3.5 | 100 | 0.02 | 66 |
| 1b | ITO/1b/Au | 477 | 589 | - | - | 0.25 | 67 |
| 1c | ITO/PVK:1c/PBD/Alq$_3$/LiF/Al | 460 | 612 | 11.0 | 500 | - | 68 |
| 1c | ITO/PVK:1c/PBD/Alq$_3$/LiF/Al | 460 | 612 | 10.0 | 500 | - | 69 |
| 1c | ITO/PVK:PBD:1c/Alq$_3$/BCP/Alq$_3$/LiF/Al | - | - | 9.0-10.0 | 3400 | - | 68 |
| 2a | ITO/2a/Al | - | 630 | - | 1000 | 1.0 | 66 |
| 2b | ITO/2b/Al | - | 630 | - | - | 0.4 | 70 |
| 2b | ITO/2b/Ga:In | 455 | 660 | 2.3 | 3500 | 1.4 | 71 |
| 2b | ITO/2b/Al | - | - | 3.0 | 1000 | 1.45 | 72 |
| 2b | ITO/PMMA:2b/Al | - | - | 2.6 | 400 | 3.0 | 72 |
| 2b | ITO/Au/2b/Ga:In | 455 | 660 | 2.3 | 5000 | - | 73 |
| 2b | ITO/PMMA:2b/Ag | - | 630 | - | - | 2.7 | 74 |
| 2b | ITO/2b/Au | 451 | 605 | - | - | 0.5 | 68 |
| 2b | ITO/Alq$_3$/2b/Ga:In | - | 630 | 2.0 | - | 6.4 | 75 |
| 2c,d | ITO/2c or 2d/Al | - | 690 | - | 600 | 0.4 | 70 |
| 2f | ITO/PMMA:2f/Ag | - | 630 | - | - | 4.1 | 74 |
| 2f | ITO/2f/Au | 459 | 610 | - | - | 0.75 | 76 |
| 2f | ITO/2f/Au | - | - | 5.0 | - | 0.6 | 76 |
| 2g | ITO/PMMA:2g/Ag | - | 630 | - | - | 5.5 | 74 |
| 2h | ITO/2h/Au | 454 | 612 | - | - | 0.5 | 76 |
| 2i | ITO/2i/Au | 454 | 609 | - | - | 0.25 | 76 |
| 3 | ITO/PVK:3/BCP/Alq$_3$/Al | 460 | 620 | 8.0 | 95 | 0.14 | 77 |
| 4a-Ru | PVA | - | 665 | 5 | 730 | 0.1 | 78 |
| 5a-Ru | PVA | - | 700 | 5 | 130 | 0.07 | 79 |
| 4b-Ru | PVA | - | 710 | 5 | 160 | 0.04 | 78 |
| 5b-Ru | PVA | - | 720 | 7 | 190 | 0.03 | 79 |
| 6 | ITO/PEDOT/PVK/PBD:6/Ba/Al | - | - | 12.5 | 930 | 0.095 | 80 |
|  |  | - | - | 14 | 240 | 0.057 | 80 |
| 7 | ITO/PEDOT/PVK/PBD:7/Ba/Al | - | - | 11.5 | 710 | 0.91 | 80 |
|  |  | - | - | 12.5 | 624 | 0.36 | 80 |
| 8 | ITO/PEDOT/PVK/PBD:8/Ba/Al | - | - | 10.5 | 2600 | 0.57 | 80 |
|  |  | - | - | 11.5 | 1970 | 0.95 | 80 |
| 9 | ITO/PEDOT/PVK/PBD:9/Ba/Al | - | - | 9 | 3670 | 0.97 | 80 |
|  |  | - | - | 7.5 | 2060 | 1.22 | 80 |
| 10 | ITO/PEDOT/PVK/PBD:10/Ba/Al | - | - | 8 | 2850 | 1.46 | 80 |
|  |  | - | - | 12.5 | 930 | 0.095 | 80 |
| 11 | Without PMMA | - | 734 | - | - | - | 81 |
|  | With PMMA | - | 734 | - | 0.32 | 0.001 | 81 |
| 12 | Without PMMA | - | 729 | - | 1.81 | 0.016 | 81 |
|  | With PMMA | - | 729 | - | 3.36 | 0.028 | 81 |
| 13 | Without PMMA | - | 744 | - | 0.75 | 0.007 | 81 |
|  | With PMMA | - | 744 | - | 0.64 | 0.013 | 81 |
| 14 | ITO/PEDOT:PSS/Ru(phen)$_3$]$^{2+}$ /Al |  | 670 | - | 120 | 0.114 | 82 |
|  | ITO/PEDOT:PSS/(BMIM-BF)- | - | 670 | - | 350 | 0.569 | 82 |